\documentclass[aps,prl,reprint,longbibliography,nobalancelastpage]{revtex4-2}

\usepackage{amsmath,amssymb}
\usepackage{graphicx}
\usepackage{xcolor}
\usepackage{mathrsfs}
\usepackage{newtxtext}
\usepackage[cmintegrals]{newtxmath}
\usepackage{stmaryrd}
\usepackage{hyperref}
\usepackage{enumitem}
\usepackage{accents}
\hypersetup{colorlinks,linkcolor={blue!80!black},citecolor={blue!80!black},urlcolor={blue!80!black}}

\renewcommand{\vec}[1]{\boldsymbol{#1}}

\renewcommand{\pi}{\uppi}

\newcommand{\mat}[1]{\mathsf{#1}}

\newcommand{\figref}[2]{[Fig.~\hyperref[#1]{\ref*{#1}(#2)}]}
\newcommand{\figrefi}[2]{[Fig.~\hyperref[#1]{\ref*{#1}(#2)}, inset]}
\newcommand{\textfigref}[2]{Fig.~\hyperref[#1]{\ref*{#1}(#2)}}

\newcommand{\figrefp}[2]{\hyperref[#1]{\ref*{#1}(#2)}}

\renewcommand{\leq}{\leqslant}
\renewcommand{\geq}{\geqslant}
\DeclareMathOperator{\tr}{tr}
\DeclareMathAlphabet{\mathcal}{OMS}{cmsy}{m}{n}

\begin{document}

\title{Turing's diffusive threshold in random reaction-diffusion systems}
\author{Pierre A. Haas}
\email{haas@pks.mpg.de}
\altaffiliation[current address: ]{Max Planck Institute for the Physics of Complex Systems, N\"othnitzer Stra\ss e 38, 01187 Dresden, Germany}
\affiliation{Mathematical Institute, University of Oxford, Woodstock Road, Oxford OX2 6GG, United Kingdom}
\author{Raymond E. Goldstein}
\email{r.e.goldstein@damtp.cam.ac.uk}
\affiliation{Department of Applied Mathematics and Theoretical Physics, Centre for Mathematical Sciences, 
\\ University of Cambridge, 
Wilberforce Road, Cambridge CB3 0WA, United Kingdom}
\date{\today}%
\begin{abstract}
Turing instabilities of reaction-diffusion systems can only arise if
the diffusivities of the chemical species are sufficiently different.  This threshold is 
unphysical in most systems with $N=2$ diffusing species, 
forcing experimental realizations of the instability to rely on fluctuations or additional 
nondiffusing species. 
Here we ask whether this diffusive threshold lowers for $N>2$ to allow ``true'' Turing 
instabilities. Inspired by May's analysis of the stability of random ecological communities, we analyze the 
probability distribution of the diffusive
threshold in reaction-diffusion systems defined by random matrices describing linearized dynamics near a homogeneous fixed point. In the numerically tractable cases $N\leq 6$, we 
find that the diffusive threshold becomes more likely to be smaller and physical as $N$ increases and that most of these many-species instabilities cannot be described by reduced models with fewer species.
\end{abstract}

\maketitle 
In 1952, Turing described the pattern-forming instability that now bears his name~\cite{turing52}: diffusion can 
destabilize a fixed point of a system of reactions that is stable in well-mixed conditions. Nigh on threescore 
and ten years on, the contribution of Turing's mechanism to chemical and biological morphogenesis remains 
debated, not least because of the \emph{diffusive threshold} inherent in the mechanism: chemical species in 
reaction systems are expected to have roughly equal diffusivities, yet Turing instabilities cannot arise 
at equal diffusivities~\cite{vastano87,pearson89}. It remains an 
open problem to determine how much of a diffusivity 
difference is required for generic 
systems to undergo this instability, yet this diffusive threshold has been 
recognized at least since reduced models of the Belousov--Zhabotinsky reaction~\cite{becker85,rovinskii87} 
only produced Turing patterns at unphysically large diffusivity differences. 

For this reason, the first experimental realizations of Turing instabilities~\cite{castets90,dekepper91,ouyang91} 
obviated the threshold by using gel reactors that greatly reduced the effective diffusivity of one species~\cite{lengyel91,dulos96}. (Analogously, biological membrane transport dynamics can increase the effective diffusivity difference~\cite{recho19}.) Later work showed
how binding to an immobile substrate, or more generally, a third, nondiffusing species, 
can allow Turing instabilities even if the $N=2$ diffusing species have equal
diffusivities~\cite{lengyel92,pearson92,korvasova15}. Such nondiffusing species continue to permeate more recent work on the network topology of Turing systems~\cite{marcon16,diego18}.

Moreover, Turing instabilities need not be deterministic: fluctuation-driven instabilities in reaction-diffusion 
systems have noise-amplifying properties that allow their pattern amplitude to be comparable to that of deterministic 
Turing patterns~\cite{biancalani17}, with a lower diffusive threshold than the deterministic one~\cite{butler09,biancalani10,butler11,stavans18}.
A synthetic bacterial population 
with $N=2$ 
species that exhibits patterns in agreement with such a stochastic Turing instability, but 
does not satisfy the conditions for a deterministic instability~\cite{karig18}, was reported recently.

These experimental instabilities relying on fluctuations or the dynamics of additional 
nondiffusing species and the nonlinear instabilities arising from finite-amplitude perturbations~\cite{vastano87} are not however instabilities in Turing's own image.
Can such instabilities be realized instead
in systems with $N>2$ diffusing species? Equivalently, is the diffusive threshold lower in 
such systems? These questions have remained unanswered, perhaps because, in contrast to the textbook 
case $N=2$ and the concomitant picture of an ``inhibitor'' out-diffusing an ``activator''~\cite{koch94,* [] [{ Vol. I, Appendix B.1, pp. 507--509 and Vol. II, Chap. 2, 
pp. 71--140, 3rd ed.}] murray}, 
the complicated instability conditions for $N>2$~\cite{satnoianu00} do not lend themselves to analytical progress.

Here, we analyze the diffusive threshold for Turing instabilities with $2\leq N\leq 6$ diffusing species. Inspired by 
May's work on the stability of random ecological communities~\cite{may72}, we analyze \emph{random 
Turing instabilities} by sampling random matrices that represent 
the linearized reaction dynamics of otherwise unspecified reaction-diffusion 
systems.  
A semianalytic approach shows that the diffusive threshold 
is more likely to be smaller and physical for $N=3$ compared to $N=2$, and that two of the 
three diffusivities are equal at the transition to instability. We extend these results to the 
remaining numerically tractable cases of reaction-diffusion systems with $4\leq N\leq 6$ 
and two 
different diffusivities: their Turing instabilities are still more likely to have a smaller and physical diffusive threshold, but most of them cannot be described by reduced models with fewer species.

We begin with the simplest case, $N=2$, in which species $u$, $v$ obey
\begin{align}
&\dot{u}=f(u,v)+d_u\nabla^2u,&&\dot{v}=g(u,v)+d_v\nabla^2v.\label{eq:rd2}
\end{align}
The conditions for Turing instability in this system~\cite{murray} only depend on the four entries 
of the Jacobian
\begin{align}
\mat{J}=\left(\begin{array}{cc}
f_u&f_v\\ g_u&g_v                 
\end{array}\right),\label{eq:J2}
\end{align}
the partial derivatives of the reaction system at a fixed 
point $(u_\ast,v_\ast)$ of the 
homogeneous system. This fixed point is stable to homogeneous perturbations iff 
${J\equiv\det{\mat{J}}>0}$ and ${I_1\equiv\tr{\mat{J}}<0}$. A stable fixed point of this kind is unstable 
to a Turing instability only if $p\equiv -f_ug_v>0$~\cite{murray}. Defining the diffusion coefficient 
ratio $D_2=\max{\{d_u/d_v,d_v/d_u\}}\geq1$, a Turing instability occurs iff these conditions hold along 
with~\footnote{See Supplemental Material at [url to be inserted], which includes Refs.~\cite{pearson89,murray,mpmath,algebra,smith18,hinch,lidskii}, for details of calculations for $N=2$, the derivation of the semianalytic approach for $N>2$ and a discussion of its numerical implementation, statistics of the wavenumber of Turing instabilities, a discussion of ``slow'' species, the proof of an asymptotic result, and for \texttt{python3} code.}\nocite{mpmath,algebra,smith18,hinch,lidskii}
\begin{align}
D_2\geq D_2^\ast\equiv\left(\dfrac{\sqrt{J\vphantom{p}}+\sqrt{J+p}}{\min{\{|f_u|,|g_v|\}}}\right)^2. \label{eq:Turing2}
\end{align}

\begin{figure}[t]
\includegraphics{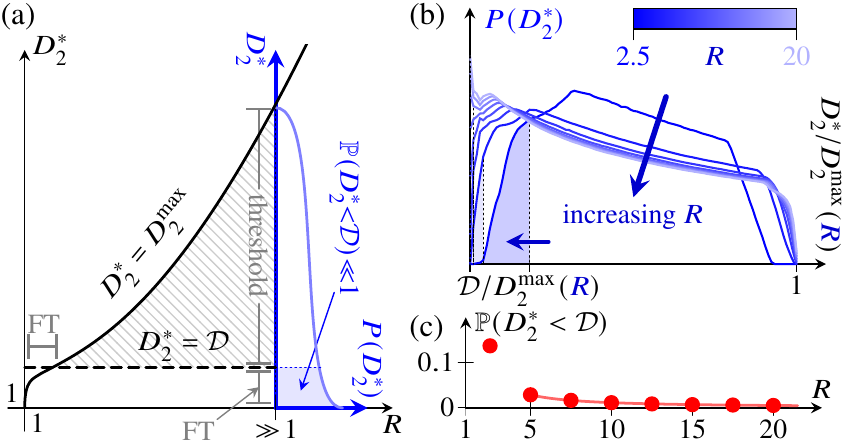} 
\caption{Turing's diffusive threshold for $N=2$. (a)~Cartoon of the diffusive threshold and the 
fine-tuning (FT) problem for $R\approx 1$ and $R\gg 1$. The diffusivity difference required mathematically is unphysical in the hatched region ${\mathcal{D}\leq D_2^\ast\leq D_2^{\max}}$. (b)~Distribution $P(D_2^\ast)$, 
supported on the (scaled) interval $[1,D_2^{\max}(R)]$, estimated for different~$R$. (c)~Plot of $\mathbb{P}(D_2^\ast<\mathcal{D})$ [shaded areas in panels (a) and (b)] against 
$R$, revealing the diffusive threshold. Markers: estimates from panel (b); solid line: exact result~\cite{Note1} for ${R>\mathcal{D}}$~\cite{fignote}.}\label{fig1}
\end{figure}

This diffusivity difference $D_2^\ast$, which is required mathematically for 
instability, is unphysical~\figref{fig1}{a} if it exceeds the diffusivity 
difference $\mathcal{D}>1$ of the physical system:
$\mathcal{D}\approx 1$ for similarly sized molecules in solution, but, e.g., 
$\mathcal{D}\approx 20$ for the stochastic Turing instability observed in 
Ref.~\cite{karig18}. Hereinafter we take $\mathcal{D}=5$ arbitrarily (but have checked that the value of $\mathcal{D}$ does not affect results qualitatively). To quantify $D_2^\ast$,
we introduce the \emph{range} $R$ of kinetic parameters,
\begin{align}
R\equiv\frac{\max{\{|f_u|,|f_v|,|g_u|,|g_v|\}}}{\min{\{|f_u|,|f_v|,|g_u|,|g_v|\}}}.
\end{align}
Equivalently, 
$f_u,f_v,g_u,g_v\in I\equiv [-R,-1]\cup[1,R]$ up to scaling, with one parameter equal to $\pm 1$ and one 
equal to $\pm R$. One deduces~\cite{Note1}
that 
\begin{align}
D_2^\ast\leq D_2^{\max}(R)\equiv\left(R+\sqrt{R^2-1}\right)^2,\label{eq:D2max}
\end{align}
as shown in \textfigref{fig1}{a}. 
Note that $D_2^{\max}\!\to\! 1$ as $R\!\to\! 1$; in this limit, 
there is no diffusive threshold: $R\approx 1$ is a particular instance of the converse \emph{fine-tuning problem} for the 
reaction kinetics that allows Turing instabilities at nearly equal diffusivities more generally~\cite{pearson89}. If $R\gg 1$, then $D_2^{\max}=O\bigl(R^2\bigr)$. This does not imply the existence of a threshold, for it does not preclude most systems with range $R$ having $D_2^\ast\ll D_2^{\max}$. The existence of a diffusive threshold therefore relates to the distribution of $D_2^\ast$ for systems with range $R$.

To understand this distribution, we draw inspiration from May's statistical analysis of the stability of ecological 
communities~\cite{may72}, which studies random Jacobians, corresponding to equilibria of otherwise unspecified 
population dynamics. By analogy, we study random Turing 
instabilities, sampling uniformly and independently random Jacobians corresponding to 
equilibria of otherwise unspecified reaction kinetics, and analyze the criteria for them to be Turing unstable. 
There is of course no more reason to expect the kinetic parameters to be independent or uniformly distributed 
than there is reason to expect the linearized population dynamics in May's analysis~\cite{may72} to 
be independent or normally distributed. Yet, in the absence of experimental understanding of what the distributions 
of these parameters should be (in either context), the potential 
of the random matrix approach to reveal stability principles has been amply demonstrated in population dynamics 
\cite{allesina12,mougi12,coyte15,grilli16,gibbs18,servan18,butler18,stone18,maynard19,haas20,barron20}.

We sample the kinetic parameters in Eq.~\eqref{eq:J2}
independently and uniformly 
from $I$, set one of them equal to $\pm 1$ and one equal to $\pm R$, and thus estimate the probability distribution $P(D_2^\ast)$ for fixed $R$~\figref{fig1}{b}. The 
threshold is 
quantified by the probability of a Turing instability being physical,
\begin{equation}
{\mathbb{P}(D_2^\ast<\mathcal{D})}=\int_1^{\mathcal{D}}{P(D_2^\ast)\,\mathrm{d}D_2^\ast}.
\end{equation}
Both from the estimates in \textfigref{fig1}{b} and by evaluating the integral in closed form~\cite{Note1}, we find that ${\mathbb{P}(D_2^\ast<\mathcal{D})}$ is tiny~\figref{fig1}{c}, except if $R$ is small, which is the fine-tuning problem again. In other words, the required diffusivity difference is very likely to be unphysical. This expresses Turing's diffusive threshold for $N=2$.

To investigate how this threshold changes with $N$ we consider 
next the $N=3$ system
\begin{subequations}\label{eq:rd3}
\begin{align}
\dot{u}&=f(u,v,w)+d_u\nabla^2u,\\
\dot{v}&=g(u,v,w)+d_v\nabla^2v,\\
\dot{w}&=h(u,v,w)+\nabla^2w,
\end{align}
\end{subequations}
where we have rescaled space to set $d_w=1$. We introduce the matrix of diffusivities and the reaction 
Jacobian,
\begin{align}
&\mat{D}=\left(\begin{array}{ccc}
d_u&0&0\\0&d_v&0\\0&0&1
\end{array}\right),&&\mat{J}=\left(\begin{array}{ccc}
f_u&f_v&f_w\\ g_u&g_v&g_w\\ h_u&h_v&h_w
\end{array}\right),\label{eq:mat3}
\end{align}
in which the entries of $\mat{J}$ are again the partial derivatives evaluated at a fixed 
point $(u_\ast,v_\ast,w_\ast)$ of the homogeneous system. This fixed point is unstable to a Turing 
instability if it is stable but, for some eigenvalue $-k^2<0$ of the Laplacian,
$\smash{\overline{\mat{J}}}\bigl(k^2\bigr)=\mat{J}-k^2\mat{D}$ is unstable~\cite{pearson89}, i.e. has an eigenvalue $\lambda$ such that $\operatorname{Re}(\lambda)<0$. More precisely, 
a Turing instability arises when a real eigenvalue of $\smash{\overline{\mat{J}}\bigl(k^2\bigr)}$ crosses zero, i.e. when
$\mathcal{J}\smash{\bigl(k^2\bigr)}\equiv\det{\smash{\overline{\mat{J}}\bigl(k^2\bigr)}}=0$, and therefore 
arises first at a wavenumber $k=k_\ast$ with $\mathcal{J}\bigl(k_\ast^2\bigr)=\partial\mathcal{J}/\partial
k^2\bigl(k_\ast^2\bigr)=0$~\cite{pearson89}. Hence $\mathcal{J}$, a cubic polynomial in $k^2$, has a 
double root at $k^2=k_\ast^2>0$, so its discriminant~\cite{algebra} vanishes. This discriminant, $\Delta(d_u,d_v)$, is a 
polynomial in $d_u,d_v$. We denote by $K(d_u,d_v)$ the double root of $\mathcal{J}$ corresponding to a point $(d_u,d_v)$ on the curve $\Delta(d_u,d_v)=0$.

\begin{figure}[b]
\includegraphics{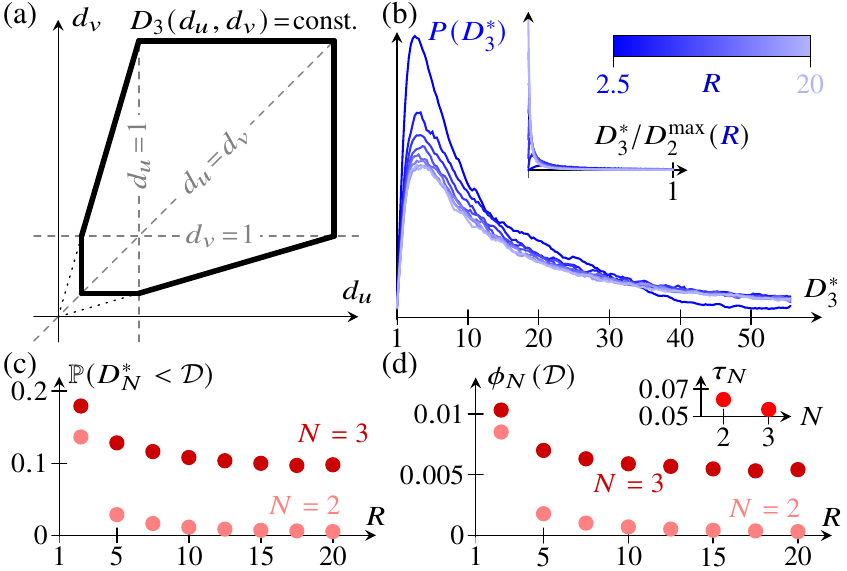} 
\caption{Results for $N=3$. (a)~Contours of $D_3(d_u,d_v)$ in the  
positive $(d_u,d_v)$ quadrant. (b)~Smoothed distribution $P(D_3^\ast)$, estimated for different $R$. Inset: same 
plot, scaled to $[1,D_2^{\max}(R)]$ for comparison to $N=2$ in \textfigref{fig1}{a}. (c)~$\mathbb{P}(D_N^\ast<\mathcal{D})$ against $R$ for ${N\in\{2,3\}}$: the diffusive 
threshold lowers for $N=3$ compared to $N=2$. (d)~Proportion $\phi_N(\mathcal{D})$ of random Jacobians 
that have a physical Turing instability, plotted against $R$, for $N\in\{2,3\}$. Inset: proportion $\tau_N$ of random Jacobians that have a (physical or unphysical) Turing instability, averaged over $R$, for $N\in\{2,3\}$~\cite{fignote}.}
\label{fig2}
\end{figure}

Determining the diffusive threshold for Turing instability in Eqs.~\eqref{eq:rd3} thus requires solving the problem
\begin{align}
\text{minimize }D_3(d_u,d_v)\quad\text{subject to}\left\{\begin{array}{ll}\Delta(d_u,d_v)=0,\\
K(d_u,d_v)>0,\end{array}\right.\hspace{-2mm}\label{eq:D3min}
\end{align}
in which the diffusion coefficient ratio is
\begin{align}
D_3(d_u,d_v)\!=\!\max{\!\{d_u,1/d_u,d_v,1/d_v,d_u/d_v,d_v/d_u\}}.\hspace{-1mm}  
\end{align}
With the aim in mind of obtaining statistics for the minimal value $D_3^\ast$, 
direct numerical solution of this constrained optimization problem is obviously not a feasible approach. In the Supplemental Material~\cite{Note1}, we therefore show how to reduce solving problem~\eqref{eq:D3min} to polynomial root finding. This semianalytic approach reveals a particular class of minima, attained at the vertices of the contours of $D_3(d_u,d_v)$~\figref{fig2}{a}, i.e. at $d_u=1$, $d_v=1$, or $d_u=d_v$. In these cases, $\Delta(d_u,d_v)=0$ is a (sextic) polynomial in the single variable $d_v$, $d_u$, or $d=d_u=d_v$, respectively. We call these minima ``binary'', since the corresponding systems have only two different diffusivities. We implement 
this approach numerically~\cite{Note1}, and sample random systems similarly to the 
case $N=2$, drawing the entries of $\mat{J}$ in Eq.~\eqref{eq:mat3} uniformly and independently 
at fixed range~$R$.

Remarkably, all global minima we found numerically were binary~\cite{Note1}. This means that the minimizing 
systems come in two flavors: those with two ``fast'' diffusers and one ``slow'' diffuser, and those 
with one ``fast'' diffuser and two ``slow'' diffusers. Systems with a nondiffusing species are a 
limit of the former; this point will be discussed below. The latter arise in models of 
scale pattern formation in fish and lizards~\cite{nakamasu09,manukyan17}, in which short-range 
pigments respectively activate and inhibit a long-range factor.

The distribution of $D_3^\ast$, shown for different values of $R$ in~\textfigref{fig2}{b}, has a different shape from that of $D_2^\ast$~[Figs.~\figrefp{fig1}{a} and \figrefp{fig2}{b}, inset]. 
While the support of the distribution of $D_3^\ast$ does not appear to be bounded, \textfigref{fig2}{c} shows that
${\mathbb{P}(D_3^\ast<\mathcal{D})>\mathbb{P}(D_2^\ast<\mathcal{D})}$. Hence the diffusivity difference is more likely to be physical for $N=3$ than for $N=2$: the diffusive threshold is lowered.

The proportion $\tau_N$ of random kinetic Jacobians that have a Turing instability 
(be it physical or unphysical) is smaller for $N=3$ than for $N=2$~\figrefi{fig2}{d}. 
This is not surprising, because a random Jacobian is less likely to correspond to a stable 
fixed point (which, we recall, is a necessary condition for Turing instability) for $N=3$ than 
for $N=2$, essentially because its entries have to satisfy more conditions 
for stability if $N=3$. It is therefore striking that the threshold is reduced sufficiently 
for $N=3$ compared to $N=2$ for the proportion $\phi_N(\mathcal{D})=\tau_N\mathbb{P}(D_N^\ast<\mathcal{D})$ 
of random Jacobians that have a physical Turing instability to be larger for $N=3$ than for 
$N=2$~\figref{fig2}{d}, even though a Turing instability of any kind is more likely if $N=2$.

To extend these results to $N>3$ diffusing species, we consider the (linearized) reaction-diffusion system
\begin{equation}
\vec{\dot{u}}=\mat{J}\cdot\vec{u}+\mat{D}\cdot\nabla^2\vec{u}, 
\end{equation}
where $\mat{J}$ is a random kinetic Jacobian, and $\mat{D}$ is a diagonal matrix of diffusivities. 
Even with our semianalytic approach, this cannot be analyzed for general 
$\mat{D}$: not even for $N=4$ were we able to obtain closed forms of the required polynomials. 
To make further progress, we therefore restrict to binary $\mat{D}$ in which the $N$ diffusivities take 
two different values only, since we showed above that $D_3^\ast$ is attained for such binary $\mat{D}$. 
As in the case $N=3$, this reduces the discriminant condition ${\Delta(\mat{D})=0}$ to polynomial equations in
one variable that determine the minimum diffusivity difference $D_N^\ast$ for Turing instability in these binary systems~\cite{Note1}.

\begin{figure}[b]
\includegraphics{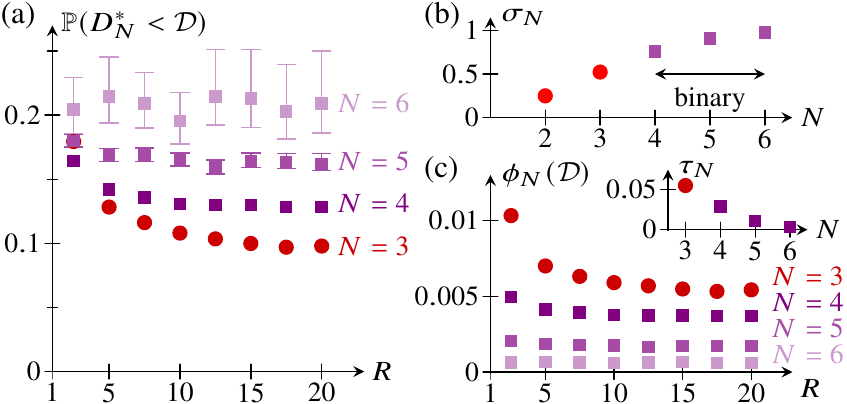}  
\caption{Results for ``binary'' systems with $4\leq N\leq 6$. 
(a)~${\mathbb{P}(D_N^\ast<\mathcal{D})}$ against $R$ for ${3\leq N\leq 6}$, revealing further lowering 
of the diffusive threshold compared to the case $N=3$. (b)~Proportion~$\sigma_N$ of random stable 
kinetic Jacobian that have a (binary, if $N>3$) Turing instability, averaged over $R$, and plotted against $N$. 
(c)~Proportion $\phi_N(\mathcal{D})$ of random Jacobians that have a physical Turing instability plotted against $R$, for $3\leq N\leq 6$. Inset: proportion $\tau_N$ of random Jacobians that have a (physical or unphysical) Turing instability, averaged over $R$, for $3\leq N\leq 6$~\cite{fignote}.}\label{fig3}
\end{figure}

Figure~\figrefp{fig3}{a} shows that the diffusive threshold lowers further for $4\leq N\leq 6$ in these systems. At the same time, the fact that most stable random kinetic Jacobians undergo such a binary Turing
instability~\figref{fig3}{b} suggests that these provide a useful picture of the diffusive threshold. However, $\tau_N$ decreases further for $N\geq 4$~\figrefi{fig3}{c}, and the widening of the bottleneck is not sufficient to prevent $\phi_N(\mathcal{D})$ from decreasing for $N\geq 4$. Nonetheless, since both $\mathbb{P}(D_N^\ast<\mathcal{D})$ and the proportion $\sigma_N$ of stable random Jacobians that are Turing unstable increase~[Figs.~\figrefp{fig3}{a} and \figrefp{fig3}{b}], so does the proportion of stable random Jacobians that have a physical Turing instability. 

How then to realize ``true'' Turing instabilities experimentally? Our analysis shows that the 
diffusive threshold of a Turing instability is more likely to be physical the more species there are, 
but how to find an experimental Turing instability in the first place? Turing instabilities 
remain rare in random reaction systems even as the number of species is increased, but the above shows 
that this rareness mainly results from the rareness of stable equilibria in such systems. The proverbial 
search for the needle in a haystack can therefore be avoided by exploring biochemical systems that 
admit a stable equilibrium, and evolving them towards a ``true'' Turing instability.

This analysis does not however reveal whether these instabilities lead to patterns that are observable 
at the physical scale of the system. Analysis of the wavenumber at which the linear instability 
first arises~\cite{Note1} suggests that we can extend our conclusions: Turing instabilities 
with more species are more likely to have physical diffusivity differences and to be observable. 
However, our statistical, linearized analysis cannot fully answer this question of observability, 
because it fundamentally depends on the system through details of the nonlinearities of its reaction 
kinetics, which set the precise nature and scale of the Turing patterns that develop beyond 
onset of the instability; this is why we have relegated this discussion to the Supplemental Material~\cite{Note1}.

\begin{figure}[t]
\includegraphics{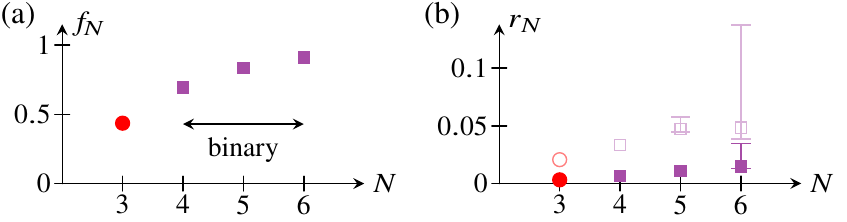}  
\caption{``Slow'' diffusers in binary Turing instabilities with $3\leq N\leq 6$. (a)~Proportion 
$f_N$ of Turing unstable systems with $n\geq 2$ ``fast'' diffusers plotted against $N$, averaged 
over~$R$. (b)~Proportion $r_N$ of systems that remain Turing unstable at $d=0$, plotted 
against $N$, averaged over~$R$. Closed markers: all Turing systems; open markers: physical Turing systems with $D_N^\ast<\mathcal{D}$~\cite{fignote}.}\label{fig4} 
\end{figure}

The different species in the systems with $3\leq N\leq 6$ analyzed above separate into ``fast'' and 
``slow'' diffusers. The diffusion of these ``slow'' species is often ignored in the analysis of systems 
of many chemical reactions~\cite{smith18}, such as the full Belousov--Zhabotinsky reaction~\cite{gyorgi}. 
Corresponding reduced models are obtained by substituting the steady-state kinetics of the 
``slow'' species into the remaining equations, thereby eliminating them from the system~\cite{smith18}. The conditions for Turing instability in these reduced models are (almost) equivalent to those for the full model with nondiffusing ``slow'' species~\cite{smith18}. However, the diffusion of the ``slow'' species cannot in general be ignored: up to reordering species and rescaling space,
\begin{align}
&\mat{D}=\left(\begin{array}{c|c}
\mat{I}&\mat{0}\\
\hline
\mat{0}&d\mat{I}
\end{array}\right),&&\mat{J}=\left(\begin{array}{c|c}
\mat{J_{11}}&\mat{J_{12}}\\\hline\mat{J_{21}}&\mat{J_{22}}               
\end{array}\right), \label{eq:fastslow}
\end{align}
where $d<1$ is the common diffusivity of the slow diffusers. Results of Ref.~\cite{smith18} imply that there is a Turing instability with nondiffusing ``slow'' species, i.e. with $d=0$, only if $\mat{J_{11}}-\mat{J_{12}}\mat{J}^{-1}_{\mat{22}}\mat{J_{21}}$ has a positive (real) eigenvalue~\cite{Note1}. Although the proportion of
Turing unstable systems that have $n\geq 2$ fast diffusers (and hence could \emph{a priori} still undergo a Turing 
instability with $d=0$) is large~\figref{fig4}{a}, the proportion of systems that do undergo such an instability is small, even if we restrict to those systems with physical diffusivity differences~\figref{fig4}{b}. Hence most of these Turing instabilities with $N>2$ species require all species to diffuse.

These many-species Turing instabilities, although binary, are thus more general than the 
instabilities of systems with nondiffusing species realized experimentally in gel reactors~\cite{castets90,dekepper91,ouyang91} and analyzed theoretically in Ref.~\cite{smith18}. 
In particular, this shows that reduced models give but an incomplete picture of Turing instabilities. 
Together with our main result, that the diffusive threshold lowers as $N$ increases, this implies 
that the failure of a reduced model to produce a physical Turing instability cannot be taken as 
an indication that a Turing instability cannot exist in the full system that the reduced model 
seeks to describe.

In this Letter, we have analyzed random Turing instabilities to show how the diffusive threshold that has 
hampered experimental efforts to generate ``true'' Turing instabilities in systems of $N=2$ diffusing species 
lowers for systems with $N\geq 3$, most of whose instabilities cannot be described by reduced models with fewer species. All of this does not, however, explain the existence of a 
``large'' threshold in the first place: even though Turing instabilities at equal diffusivities 
are impossible~\cite{vastano87,pearson89}, this does not mean that the threshold needs to be
``large''. In this context, we prove an asymptotic result in the Supplemental Material~\cite{Note1}: for a Jacobian $\mat{J}$ to allow a Turing instability at almost equal diffusivities $\mat{D}\approx\mat{I}$, $\mat{J}$ must be even closer to a singular matrix $\mat{J_0}$, i.e. $\mat{J}-\mat{J_0}\ll\mat{D}-\mat{I}$. In this sense, the threshold $\mat{D}-\mat{I}$ is asymptotically ``large''. Understanding how a large threshold 
arises more generally outside this asymptotic regime and lowers as $N$ increases remains an open problem, as do extending the present analysis to include the nonlocal interactions~\cite{ohta,interface} that arise for example in vegetation patterns~\cite{meron} and extending previous work~\cite{diego18,scholes19} on the robustness of Turing patterns to $N\geq 3$. The latter in particular may help to identify those chemical or 
biological pattern forming systems with $N\geq 3$ in which the ``true'' Turing instabilities discussed 
here can be realized experimentally.

\begin{acknowledgments}
We thank N. Goldenfeld, A. Krause, and P. K. Maini for discussions.  This work was supported in part by 
a Nevile Research Fellowship from Magdalene College, Cambridge, and a Hooke Research Fellowship (P.A.H.),
Established Career Fellowship EP/M017982/1 from the Engineering and Physical Sciences Research Council 
and Grant 7523 from the Marine Microbiology Initiative of the Gordon and Betty Moore Foundation (R.E.G.).
\end{acknowledgments}

\bibliography{turing}

\begin{thebibliography}{51}
\bibitem [{fig()}]{fignote}%
  \BibitemOpen
  \href@noop {} {}\bibinfo {note} {All results shown are for $\mathcal{D}=5$.
  The asymmetric error bars in Figs.~\ref{fig3} and \ref{fig4} show $95\%$
  confidence intervals larger than the plot markers, corrected for systems for
  which the numerics failed.}\BibitemShut {Stop}%
\end{thebibliography}
\end{document}


\title{Turing's diffusive threshold in random reaction-diffusion systems\\ --- \textsc{Supplemental Material} ---}
\author{Pierre A. Haas}
\affiliation{Mathematical Institute, University of Oxford, Woodstock Road, Oxford OX2 6GG, United Kingdom}
\author{Raymond E. Goldstein}
\affiliation{Department of Applied Mathematics and Theoretical Physics, Centre for Mathematical Sciences, \\ University of Cambridge, 
Wilberforce Road, Cambridge CB3 0WA, United Kingdom}
\maketitle
\setcounter{figure}{0}
\renewcommand{\thefigure}{S\arabic{figure}}
\setcounter{table}{0}
\renewcommand{\thetable}{S\arabic{table}}
\setcounter{equation}{0}
\renewcommand{\theequation}{S\arabic{equation}}

This Supplemental Material is divided into five sections, which provide (i) details of calculations for $N=2$, (ii) the derivation of the semianalytic approach for $N=3$ and a discussion of its numerical implementation, (iii) an analysis of the statistics of the wavenumber at which a Turing instability first arises, (iv) a discussion of Turing instabilities with nondiffusing ``slow'' species, and (v) a proof of the asymptotic result claimed in the conclusion of our Letter.

\section{Details of Calculations for $\vec{N=2}$}
\subsection{Derivation of Eq.~(\ref{eq:Turing2})}
The form of the condition for Turing instability in Eq.~\eqref{eq:Turing2} follows from that in Eq.~(2.26) on page 85 of Vol. II of 
Ref.~\cite{* [] [{ Vol. II, Chap. 2, 
pp. 71--140, 3rd ed.}] murray} which, in our notation, reads 
\begin{align}
f_u+dg_v\geq 2\sqrt{dJ},\label{eq:murray}
\end{align}
a quadratic in $d=d_u/d_v$. Hence
\begin{align}
\sqrt{d}\gtrless\sqrt{d_\ast}\equiv\dfrac{\sqrt{J}\pm\sqrt{J-\smash{f_ug_v}}}{g_v}\quad\text{if }g_v\gtrless 0.
\end{align}
We notice that Eq.~\eqref{eq:murray} requires $f_u+d g_v\geq 0$. Since $I_1<0$, this implies that $d\gtrless 1$ if $g_v\gtrless 0$. Hence $D_2^\ast=d_\ast$ if $g_v>0$, but $D_2^\ast=1/d_\ast$ if $g_v<0$. Now, if $g_v\gtrless0$, then $|f_u|\gtrless|g_v|$ because $I_1<0$ and $p>0$. Equation~\eqref{eq:Turing2} then follows, since
\begin{align}
\dfrac{g_v}{\sqrt{J\vphantom{p}}-\sqrt{J+p}}=\dfrac{\sqrt{J\vphantom{p}}+\sqrt{J+p}}{f_u}. 
\end{align}
\subsection{Derivation of Eq.~(\ref{eq:D2max})}
Equation~\eqref{eq:Turing2} shows that $D_2^\ast$ is continuous on $I^4$, so attains its maximum value on that domain. Since $p>0$ and 
$J>0$, $q\equiv -f_vg_u>0$, so that $J+p=q$. Now $D_2^\ast$ only depends on $f_v,g_u$ through $q$, and, by direct computation from Eq.~\eqref{eq:Turing2},
\begin{align}
\dfrac{\partial D_2^\ast}{\partial q}=\dfrac{D_2^\ast}{\sqrt{J(J+p)}}>0.
\end{align}
Hence $D_2^\ast$ increases with $q$, so $(f_v,g_u)=\pm(R,-R)$ at the maximum. 

Now assume that $|f_u|\geq|g_v|$. Since $I_1<0$ and $|f_u|\geq|g_v|$, it follows that $f_u<0$ and $g_v>0$. Then
\begin{align}
\dfrac{\partial D_2^\ast}{\partial f_u}&=\dfrac{\sqrt{J\vphantom{p}}+\sqrt{q\vphantom{J}}}{g_v\sqrt{J\vphantom{q}}}>0,&&\dfrac{\partial D_2^\ast}{\partial g_v}=-\dfrac{\left(\sqrt{J\vphantom{p}}+\sqrt{q\vphantom{J}}\right)^3}{g_v^3\sqrt{J\vphantom{q}}}<0,\label{eq:dD2dgv}
\end{align}
and so $(f_u,g_v)=(1,-1)$ at the maximum. If $|f_u|\leq |g_v|$, we similarly find that $(f_u,g_v)=(-1,1)$ at the maximum. Substituting these values into Eq.~\eqref{eq:Turing2} yields Eq.~\eqref{eq:D2max}.

\subsection{Calculation of $\vec{\pmb{\mathbb{P}}(D_2^\ast<\mathcalbf{D})}$ for $\vec{\mathcalbf{D}\leq R}$}
There are $48$ ways of assigning values $\pm 1$ and $\pm R$ to two of the entries $f_u,f_v,g_u,g_v$ of $\mat{J}$. Integrating the conditions for Turing instability of the remaining entries in each of these cases using \textsc{Mathematica} (Wolfram, Inc.) gives the area of parameter space in which a Turing instability arises,
\begin{align}
\iiiint_{I^4}{\vmathbb{1}\left(\begin{array}{c}J>0,\;I_1<0\\p>0\\\max{|\mat{J}|}=R\\\min{|\mat{J}|}=1\end{array}\right)\,\mathrm{d}\mat{J}}=12(R-1)^2,\label{eq:pT} 
\end{align}
where we use the shorthand $\mathrm{d}\mat{J}=\mathrm{d}f_u\,\mathrm{d}f_v\,\mathrm{d}g_u\,\mathrm{d}g_v$. 
To analyze the condition $D_2^\ast<R$, we note that the expression for $D_2^\ast$ in Eq.~\eqref{eq:Turing2} shows that we may swap $f_u,g_v$ and $f_v,g_u$. Hence the $48$ cases reduce to $4$ cases (corresponding to the entries $\pm 1$ or $\pm R$ being on the the same or on different diagonals):
\begin{align*}
&\text{(1)\quad$|f_u|=R$, $|g_v|=1$;}&&\text{(2)\quad$|f_v|=R$, $|g_u|=1$};\\
&\text{(3)\quad$|f_u|=R$, $|f_v|=1$;}&&\text{(4)\quad$|f_u|=1$, $|f_v|=R$}.
\end{align*}
Moreover, since $q>0$, we may take $f_v>0$ and $g_u<0$ without loss of generality. We now discuss these cases separately.
\begin{enumerate}[label={(\arabic{enumi})},leftmargin=*]
\item $I_1<0$ implies ${f_u=-R}$, $g_v=1$, and so
\begin{align}
D_2^\ast=\left(\sqrt{q\vphantom{R}}+\sqrt{q-R}\right)^2\geq R. 
\end{align}
\item $f_ug_v=-R$ since $q>0$, so $J=f_ug_v+R$.
\item $f_u=-R$ because $I_1<0$. Now $p,q>0$, and so ${0<J=-R|g_v|-|g_u|<0}$. This is a contradiction. 
\item $f_u=1$ as $I_1<0$. Since $g_v\leq -1$, it follows that
\begin{align}
D_2^\ast=\left(\sqrt{-g_uR}+\sqrt{-g_uR-g_v}\right)^2\geq R. 
\end{align}
\end{enumerate}
In this way, $D_2^\ast<R$ quantifies the diffusive threshold in a natural way. In particular, $D_2^\ast<R$ is only possible in case (2). Since $J>0$, we require $f_ug_v+R>0$ in that case. Now $I_1<0$ and $p>0$, so $1<f_u<-R/g_v$ or $1<g_v<-R/f_u$ depending on $f_u>0,g_v<0$ or $f_u<0,g_v>0$. Assume without loss of generality that ${|f_u|\geq|g_v|}$. Then $f_u<0$, $g_v>0$ as $I_1<0$. Moreover, using Eq.~\eqref{eq:Turing2}, $D_2^\ast=R$ if and only if $g_v=2+f_u/R$. From Eqs.~\eqref{eq:dD2dgv}, $D_2^\ast$ decreases as $g_v$ increases. Hence
\begin{align}
D_2^\ast<R\;\Longleftrightarrow\;2+f_u/R<g_v\leq-R/f_u\text{ and }f_u+g_v<0, \label{eq:leqRcond}
\end{align}
using the conditions derived previously. Note that $-R/f_u<R$ and $2+f_u/R>1$ for $-R<f_u<-1$. If $|f_u|<|g_v|$, $f_u,g_v$ are swapped in these conditions. Moreover, since $q>0$, case (2) corresponds to $4$ of the $48$ cases. Hence we obtain, again using \textsc{Mathematica},
\begin{align}
\iiiint_{I^4}{\vmathbb{1}\left(\begin{array}{c}J>0,\;I_1<0\\p>0,\;D_2^\ast<R\\\max{|\mat{J}|}=R\\\min{|\mat{J}|}=1\end{array}\right)\;\mathrm{d}\mat{J}}=4\left(\dfrac{2R(1-R)}{1+R}+R\log{R}\right).\label{eq:pD2}
\end{align}
Equations~\eqref{eq:pT} and \eqref{eq:pD2} imply
\begin{align}
\mathbb{P}\left(D_2^\ast< R\right)=\dfrac{R\left[(R+1)\log{R}-2(R-1)\right]}{3(R-1)^2(R+1)}. 
\end{align}
In particular, $\mathbb{P}\left(D_2^\ast<R\right)=O(\log{R}/R)\ll 1$ for $R\gg 1$. This statement expresses the existence of the diffusive bottleneck mathematically.

From a more physical point of view, as discussed in our Letter, it is more natural to consider the probability $\mathbb{P}\left(D_2^\ast<\mathcal{D}\right)$, for some constant~$\mathcal{D}>1$. Since ``small'' values $R\leq \mathcal{D}$ require fine-tuning of the reaction kinetics, we restrict to $\mathcal{D}\leq R$, so that $D_2^\ast<\mathcal{D}$ is only possible in case (2) above. We consider again the case $g_v>0$, $f_u<0$. Similarly to the derivation of conditions~\eqref{eq:leqRcond}, we find
\begin{align}
D_2^\ast<\mathcal{D}\quad\Longleftrightarrow\quad&g_v>\sqrt{\dfrac{R}{\mathcal{D}}},\quad{-\dfrac{R}{g_v}}\leq f_u<\mathcal{D} g_v-2\sqrt{\mathcal{D}R}\nonumber\\
&\quad\text{and }f_u+g_v<0. \label{eq:leqccond}
\end{align}
In particular,
\begin{subequations}
\begin{align}
-\dfrac{R}{g_v}\!=\!\max{\left\{-R,-\dfrac{R}{g_v}\right\}}\leq f_u<\min{\!\left\{-1,-g_v,\mathcal{D} g_v\!-\!2\sqrt{\mathcal{D}R}\right\}}, 
\end{align}
in which, since $g_v\geq 1$,
\begin{align}
\min{\!\left\{-1,-g_v,\mathcal{D}g_v\!-\!2\sqrt{\mathcal{D}R}\right\}}\!=\!\left\{\begin{array}{cl}
-g_v&\hspace{-2mm}\text{if }g_v\!>\!\dfrac{2\sqrt{\mathcal{D}R}}{\mathcal{D}+1};\\
\mathcal{D}g_v\!-\!2\sqrt{\mathcal{D}R}&\text{otherwise}.
\end{array}\right.
\end{align}
\end{subequations}
We notice that $\sqrt{R}>2\sqrt{\mathcal{D}R}/(\mathcal{D}+1)\!>\!\sqrt{R/\mathcal{D}}$ since $\mathcal{D}>1$, and also that $\smash{\mathcal{D}g_v-2\sqrt{\mathcal{D}R}>-R/g_v\Longleftrightarrow\bigl(\sqrt{\mathcal{D}}g_v-\sqrt{R}\bigr)^2>0}$, but $-g_v/-R/g_v\Longleftrightarrow g_v<\sqrt{R}$. The area of parameter space described by conditions~\eqref{eq:leqccond} is therefore
\begin{subequations}
\begin{align}
&\int_{\frac{2\sqrt{\mathcal{D}R}}{\mathcal{D}+1}}^{\sqrt{R}}{\left(\int_{-\frac{R}{g_{\scalebox{0.6}[0.4]{$v$}}}}^{-g_v}{\mathrm{d}f_u}\right)\mathrm{d}g_v}+ \int_{\sqrt{\frac{R}{\mathcal{D}}}}^{\frac{2\sqrt{\mathcal{D}R}}{\mathcal{D}+1}}{\left(\int_{-\frac{R}{g_{\scalebox{0.6}[0.4]{$v$}}}}^{\mathcal{D}g_v-2\sqrt{\mathcal{D}R}}{\mathrm{d}f_u}\right)\mathrm{d}g_v}\nonumber\\
&\quad=\dfrac{R}{2}\log{\mathcal{D}}-\dfrac{\mathcal{D}-1}{\mathcal{D}+1}R.
\end{align}
Hence~\footnote{We were not able to compute this integral using \textsc{Mathematica}. This is why, by contrast with the derivation of Eq.~\eqref{eq:pD2}, we have given the full details of the calculation above.}
\begin{align}
\iiiint_{I^4}{\vmathbb{1}\left(\begin{array}{c}J>0,\;I_1<0\\p>0,\;D_2^\ast<\mathcal{D}\\\max{|\mat{J}|}=R\\\min{|\mat{J}|}=1\end{array}\right)\;\mathrm{d}\mat{J}}=4\left[2\left(\dfrac{R}{2}\log{\mathcal{D}}-\dfrac{\mathcal{D}-1}{\mathcal{D}+1}R\right)\right],\label{eq:pD2c}
\end{align}
\end{subequations}
for $R>\mathcal{D}$, and, as above, we conclude that, for $R>\mathcal{D}$,
\begin{align}
\mathbb{P}\left(D_2^\ast<\mathcal{D}\right)=\dfrac{R}{3(R-1)^2}\left[\log{\mathcal{D}}-\dfrac{2(\mathcal{D}-1)}{\mathcal{D}+1}\right].  
\end{align}

\subsection{Nondimensionalization}
We close by remarking on the (absence of) nondimensionalization of the reaction system. Indeed, up to rescaling time, 
one among $f_u,f_v,g_u,g_v$ can be set equal to $\pm 1$. Moreover, one more parameter can be set equal to $\pm 1$ 
by rescaling $u,v$ differently. However, if we made those choices, we could no longer sample from a fixed interval. 

\section{Semianalytic method for $\vec{N=3}$}
\subsection{Derivation of the semianalytic method}
\subsubsection{Preliminary observations}
Before deriving the semianalytic method, we need to make two preliminary observations. 

First, the necessary and sufficient (Routh--Hurwitz) conditions for the homogeneous system to be 
stable include $I_1\equiv\tr{\mat{J}}<0$ and ${J\equiv\det{\mat{J}}<0}$~\cite{murray}. By definition,
\smash{$\overline{\mat{J}}\bigl(k_\ast^2\bigr)$} has one zero eigenvalue. The other two eigenvalues are either real 
or two complex conjugates $\lambda,\lambda^\ast$. In the second case, they are both stable (i.e. have negative real parts) since
\begin{align}
2\,\text{Re}(\lambda)=0+\lambda+\lambda^\ast=\tr{\overline{\mat{J}}\bigl(k_\ast^2\bigr)}=I_1-k_\ast^2\tr{\mat{D}}<I_1<0. 
\end{align}
Hence Eqs.~\eqref{eq:rd3} are not unstable to an oscillatory (Turing--Hopf) 
instability at $(d_u^\ast,d_v^\ast)$, so, by minimality of $(d_u^\ast,d_v^\ast)$, the system destabilizes to a Turing 
instability there. 

Moreover, since $\mathcal{J}$, viewed as a polynomial in $k_\ast^2$, has leading coefficient $-d_ud_v$ and constant term $\mathcal{J}(0)=J<0$, the double 
root $K(d_u,d_v)$ varies continuously with $d_u,d_v$ and cannot change sign on a branch of $\upDelta(d_u,d_v)=0$ in the 
positive $(d_u,d_v)$ quadrant.

\subsubsection{Reduction of problem~(\ref{eq:D3min}) to polynomial equations}
The discriminant of $\mathcal{J}$, viewed as a polynomial in the two variables $d_u,d_v$, is
\begin{align}
\Delta(d_u,d_v)=\sum_{m=0}^4{\sum_{n=0}^4{\delta_{mn}d_u^md_v^n}},\label{eq:D3}
\end{align}
where $\delta_{00}=\delta_{10}=\delta_{01}=\delta_{34}=\delta_{43}=\delta_{44}=0$ and (complicated) expressions for the 19 non-zero coefficients can be found in terms of the entries of $\mat{J}$ using \textsc{Mathematica} (Wolfram, Inc.). 

The second remark above implies that, at a local minimum of $D_3(d_u,d_v)$ on $\Delta(d_u,d_v)=0$, one of the 
following occurs: 
\begin{enumerate}[label={(\roman{enumi})},leftmargin=*,widest=iii]
 \item $\Delta(d_u,d_v)=0$ is tangent to a contour of $D_3(d_u,d_v)$;
 \item $\Delta(d_u,d_v)$ intersects a vertex of a contour of $D_3(d_u,d_v)$;
 \item $\Delta(d_u,d_v)$ is singular.
\end{enumerate}
The contours of $D_3(d_u,d_v)$ are drawn in~\textfigref{fig2}{a} of our Letter and show that tangency to a contour in case (i) 
requires
\begin{align}
\mathrm{d}d_u=0\quad\text{or}\quad\mathrm{d}d_v=0\quad\text{or}\quad\mathrm{d}d_v/\mathrm{d}d_u=d_v/d_u. 
\end{align}
Since $\Delta(d_u,d_v)=0$, the chain rule reads 
\begin{align}
0=\mathrm{d}\Delta=\dfrac{\partial\Delta}{\partial d_u}\,\mathrm{d}d_u+\dfrac{\partial\Delta}{\partial d_v}\,\mathrm{d}d_v.
\end{align}
Hence there are two subcases:
\begin{enumerate}[label={(\alph{enumi})},leftmargin=*,widest=b]
\item $\dfrac{\partial\Delta}{\partial d_v}=0$ or $\dfrac{\partial\Delta}{\partial d_u}=0$;
\item $d_u\dfrac{\partial\Delta}{\partial d_u}+d_v\dfrac{\partial\Delta}{\partial d_v}=0$.
\end{enumerate}
In subcase (a), $\Delta$ viewed as a polynomial in $d_v$ or $d_u$ 
has a double root, and so its discriminant~\cite{algebra} must vanish. On removing zero roots, this 
discriminant of a discriminant is found to be a polynomial of degree $20$ in $d_u$ or $d_v$, 
respectively; complicated expressions for its coefficients in terms of the non-zero coefficients 
$\delta_{mn}$ in Eq.~\eqref{eq:D3} are obtained using \textsc{Mathematica}. Similarly, in subcase 
(b), the resultant~\cite{algebra} of $\Delta$ and $d_u\partial\Delta/\partial d_u+d_v\partial\Delta/\partial d_v$, 
viewed as polynomials in $d_u$ or $d_v$ must vanish. This resultant is another polynomial of degree $20$ in 
$d_v$ or $d_u$. 

Next, in case (ii), $d_u=1$ or $d_v=1$ or $d_u=d_v$~\figref{fig2}{a}, which reduces $\Delta$ 
to three different polynomials in the single variable $d_v$, $d_u$, or $d=d_u=d_v$, respectively. These polynomials have degree $6$.

Finally, in case (iii), we note that, at a singular point, ${\Delta=\partial\Delta/\partial d_u=\partial\Delta/\partial d_v=0}$, 
and so we are back in case~(i), subcase (a). 

Thus, we have reduced finding candidates for local minima 
in~\eqref{eq:D3min} to solving polynomial equations: this defines our semianalytic approach. The global minimum is found 
among those local minima with $K(d_u,d_v)>0$; in case~(i), the roots only correspond to local 
minima if additionally $d_u,d_v>1$ or $d_u,d_v<1$ in subcase (a) and $d_u<1<d_v$ or $d_v<1<d_u$ in 
subcase (b)~\figref{fig2}{a}.
\subsubsection{Extension to binary systems with $N>3$}
For binary systems, the diagonal entries of $\mat{D}$ take two different values, $d_1,d_2$ only. Up to rescaling space, $d_1=1$ and $d_2=d$, which turns the condition $\Delta(\mat{D})=0$ into $2^{N-1}-1$ different polynomial equations in the single variable $d$, corresponding to the different combinatorial ways of assigning diffusivities $d_1,d_2$ to the $N$ species (in such a way that not all species have the same diffusivity). Determining the minimum value $D_N^\ast$ of $D_N=\max{\{d,1/d\}}$ for these binary systems is thus reduced, again, to solving polynomial equations.

The argument we used above to show that coexistence of Turing and Turing--Hopf instabilities is not possible for $N=3$ does not, however, carry over to $N>3$. Numerically, it turns out, however, that systems in which Turing and Turing--Hopf instabilities coexist are rare. We therefore treat these systems in the same way as we treat systems for which the numerics fail (as discussed below).
\subsection{Numerical implementation}
Implementing the semi-analytical approach for $N=3$ and its extension to binary systems with $4\leq N\leq 6$ numerically takes some care as the coefficients of the polynomials that arise can range over many orders of magnitude. Our \texttt{python3} implementation therefore uses the \texttt{mpmath} library for variable precision arithmetic~\cite{mpmath}.

To determine the positive real roots of the polynomials that arise in the semi-analytical approach, we complement the Durand--Kerner complex root finding implemented in the \texttt{mpmath} library~\cite{mpmath} with a test based on Sturm's theorem~\cite{algebra}, to ensure that all positive real roots are found. Those systems in which root finding fails---either because the Durand--Kerner algorithm fails to converge or because it finds an incorrect number of positive real roots---are discarded, but included in error estimates where reported.

\subsection{Numerical samples}
Table~\ref{tab1} gives the number of random Turing unstable systems from which distributions, averages, and probabilities were estimated for each ${R\!\in\!\{2.5,5,7.5,10,12.5,15,17.5,20\}}$.

\begin{table}[b]
\caption{Number of random Turing unstable systems used to estimate distributions, averages, and probabilities for the different values of $N$, and corresponding figures.} \label{tab1}
\begin{ruledtabular}
\begin{tabular}{cccc}
$N$&Type&$\max{T}$\footnote{Maximal number of Turing unstable systems.}&Reference\footnote{Figure (if any) in which results are shown.}\\
\hline 
$N=2$&non-binary&$10^7$&Figs.~\ref{fig1}, \ref{figS1}\\
$N=3$&non-binary&$10^4$&\\
$N=3$&binary& $10^5$&Figs.~\ref{fig2}, \ref{fig4}, \ref{figS1}\\
$N=4$&binary& $10^5$&Figs.~\ref{fig3}, \ref{fig4}, \ref{figS1}\\
$N=5$&binary& $2\cdot 10^4$&Figs.~\ref{fig3}, \ref{fig4}, \ref{figS1}\\
$N=6$&binary& $2\cdot 10^3$&Figs.~\ref{fig3}, \ref{fig4}, \ref{figS1}\\
\end{tabular}
\end{ruledtabular}
\end{table}

For $N=3$, we ran both a search for general, non-binary systems and a (larger but numerically less expensive) search for binary systems only. Since the first search only yielded binary global minima (as stated in our Letter), we used the results of the second, larger search for Figs.~\ref{fig3} and \ref{fig4}.

\begin{figure}[b]
\includegraphics{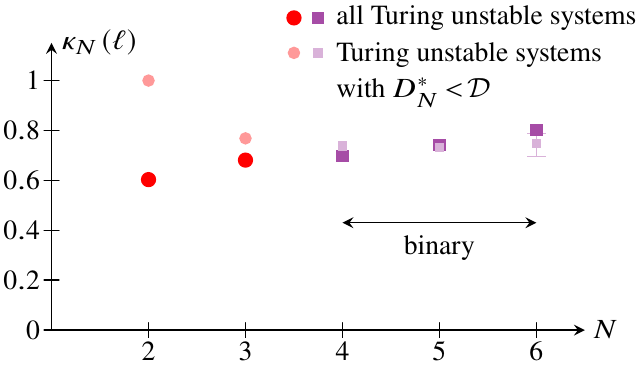}
\caption{Wavenumber statistics. Probability $\kappa_N(\ell)$ of a Turing instability being ``observable'' at a scale difference $\ell$ plotted against $N$; see text for further explanation. Larger markers: $\kappa_N(\ell)$ estimated from all Turing unstable systems; smaller markers: $\kappa_N(\ell)$ estimated from only those Turing unstable systems with $\smash{D_{\smash N}^\ast< \mathcal{D}}$. Parameter values: $R=10$, $\ell=10$, $\mathcal{D}=5$. Asymmetric error bars again correspond to $95\%$ confidence intervals larger than the plot markers, corrected for systems for which the numerics failed.}\label{figS1}
\end{figure}

\section{Wavenumber Statistics}
In this Section, we discuss the wavenumber $k_N^\ast$ at which a Turing instability first arises at $D_N=D_N^\ast$. In particular, as discussed in our Letter, we must ask whether a Turing instability is ``observable at the system size''. This observability requires the lengthscale $1/k_N^\ast$ of the linear instability to be (a) smaller than the system $L$ and (b) larger than $L/\ell$, for some scale difference $\ell>1$. We are thus led to consider the probability $\mathbb{P}(K<k_N^\ast<\ell K)$, where $K=1/L$.

It is instructive to start by considering the case $N=2$. For the reaction-diffusion system in Eq.~\eqref{eq:rd2}, a Turing instability arises for $D_2=D_2^\ast$ at a wavenumber $k_2^\ast=(J/d_ud_v)^{1/4}$~\cite{murray}. We stress that this value depends on $d_u,d_v$ not only through their ratio $d=d_u/d_v$. To absorb the dependence on the dimensional system scale, it is natural to consider
\begin{subequations}
\begin{align}
\kappa_2(\ell)=\max_{K}{\left\{\mathbb{P}\left(K<k_2^\ast<\ell K\right)\right\}},  
\end{align}
as the maximal probability of a Turing instability being observable at some inverse system scale $K$ over a fixed scale difference $\ell$. We denote by $K_2(\ell)$ the corresponding maximizing inverse system size.

For $N>2$, we correspondingly ask: what is the probability of a Turing instability being observable at this inverse system size? We therefore define
\begin{align}
\kappa_N(\ell)= \mathbb{P}\left(K_2(\ell)<k_N^\ast<\ell K_2(\ell)\right)\quad\text{for }N>2.
\end{align}
\end{subequations}
Figure~\ref{figS1} plots $\kappa_N(\ell)$ against $N$, for fixed values of $R$ and $\ell$, but the qualitative behaviour is independent of $R$ and $\ell$. We notice that $\kappa_N(\ell)$ increases slightly with $N$. If we restrict the analysis to those Turing unstable systems with $D_N^\ast\leq\mathcal{D}$, the probability is reduced somewhat for $N>2$ compared to the case $N=2$. This merely reflects the ``fine-tuning problem'': the wavenumber is strongly constrained for those very rare systems that have a ``small'' diffusive threshold at $N=2$. Moreover, about three quarters of the Turing instabilities at $N>2$ do arise at physical wavenumbers, so we can extend the observations in   Figs.~\figrefp{fig2}{d} and \figrefp{fig3}{c} to note that random kinetic Jacobians are still more likely to be unstable to an observable Turing instability with small diffusive threshold for $N>2$ than for $N=2$.

\section{Diffusion of ``slow'' species}
In the notation of Eq.~\eqref{eq:fastslow} of our Letter, Ref.~\cite{smith18} shows that Turing instability at $d=0$ requires $\mat{J_{22}}$ to be stable (i.e. all its eigenvalues to have negative real part): if it is not, instabilities arise at arbitrarily small and therefore unphysical lengthscales. In particular, $\det{\mat{J_{22}}}\not=0$, and so, using another result of Ref.~\cite{smith18},
\begin{align}
\det{\bigl(\mat{J}-k^2\mat{D}\bigr)}=\det{\mat{J_{22}}}\det{\bigl(\mat{j}-k^2\mat{I}\bigr)}, 
\end{align}
where $\mat{j}=\mat{J_{11}}-\mat{J_{12}}\mat{J}^{-1}_{\mat{22}}\mat{J_{21}}$. Hence a Turing instability occurs at $d=0$ only if $\mat{j}$ has a positive real eigenvalue, as claimed in our Letter.

\section{The asymptotic diffusive threshold}
Let $\mat{J}=O(1)$ be a Turing unstable kinetic Jacobian, 
with an eigenvalue $\lambda$ destabilising at nearly equal diffusivities, so that ${\mat{D}=\mat{I}+\mat{d}}$ with
$\mat{d}=o(1)$. The following claim extends an argument of Ref.~\cite{pearson89}:
\begin{claim}
$\mat{J}$ has a defective zero eigenspace. 
\end{claim}
\begin{proof}
Because $\mat{J}-k^2\mat{I}$ has a stable eigenvalue $\lambda-k^2$ 
and $-k^2\mat{d}\ll\mat{J}-k^2\mat{I}$, the corresponding eigenvalue of
\begin{align*}
\mat{J}-k^2\mat{D}=\bigl(\mat{J}-k^2\mat{I}\bigr)-k^2\mat{d} 
\end{align*}
can only have positive real part if 
$\lambda-k^2=o(1)$ i.e. if $\lambda=o(1)$ and $k^2=o(1)$ since $\text{Re}(\lambda)<0$. Hence 
$\mat{J}$ and $\mat{J}-k^2\mat{I}$ have a zero eigenvalue at leading order. Additionally, the eigenvalue 
correction from $-k^2\mat{d}=o\bigl(k^2\bigr)$ must be $\smash{O\bigl(k^2\bigr)}$ at least, which occurs iff the
(leading-order) zero eigenspaces of $\smash{\mat{J}-k^2\mat{I}}$ and $\mat{J}$ are defective~\cite{hinch}; this final implication is discussed in more detail in Ref.~\cite{lidskii}. 
\end{proof}

The generic case is therefore $\mat{J}=\mat{J_0}+O(\varepsilon)$,
where $\varepsilon\ll 1$ and $\mat{J_0}$ has a defective double zero eigenvalue.
\begin{claim}
$\mat{d}\gtrsim O\bigl(\sqrt{\varepsilon}\bigr)$; in particular, $\mat{D}-\mat{I}\gg\mat{J}-\mat{J_0}$. 
\end{claim}
\begin{proof}
Since $\mat{J_0}$ has a defective double zero eigenvalue, $\mat{J}$ has two
$O(\sqrt{\varepsilon})$ eigenvalues~\cite{hinch}, assumed to be stable (i.e. to have negative real parts). With $k=O(\varepsilon^\kappa)$,
$\mat{d}=O\bigl(\varepsilon^\delta\bigr)$, destabilizing one of these requires, using the proof of the first claim above, 
$-k^2\mat{d}\gtrsim O(\varepsilon)$ and $-k^2\mat{I}\lesssim O(\sqrt{\varepsilon})$, i.e. 
$2\kappa+\delta\leq 1$ and $\kappa\geq 1/4$. Hence $\delta\leq 1/2$. This proves the claim.
\end{proof}
\section*{Supplemental Code}
The online Supplemental Material also includes excerpts from the \texttt{python3} code that we have written to implement the semianalytic approach for $N\geq 3$.
\section*{Supplemental References}
\bibliography{turing}